\documentclass[aps,pre,twocolumn,superscriptaddress]{revtex4-2}

\usepackage{graphicx}
\usepackage{xcolor}
\usepackage{dcolumn}
\usepackage{bm}
\usepackage{amsmath}
\usepackage{amssymb}

\begin{document}

\title{Dynamics due to competitive flip cycles in active Potts models}

\author{Hiroshi Noguchi}
\email[]{noguchi@issp.u-tokyo.ac.jp}
\affiliation{Institute for Solid State Physics, University of Tokyo, Kashiwa, Chiba 277-8581, Japan}
\date{\today}

\begin{abstract}
Nonequilibrium  spatiotemporal patterns have been extensively studied.
However, a single oscillator or cyclic loop of states is typically employed at each site in theories and simulations.
Here, we investigate how competition among multiple identical cyclic loops at each site alters patterns.
We simulate active Potts models
with standard Potts interactions between neighboring sites in two-dimensional square lattices.
When multiple three-state cycles exist in state flips, such as in octahedral and square-antiprism networks,
all types of spiral waves comprising the three states are formed simultaneously at high flip energies.
However, at lower energies, only one or a few types emerge and switch stochastically into different types.
At even lower energies, cyclic changes in single-state dominant homogeneous phases emerge [homogeneous cycling (HC) mode].
At intermediate flip energies,
the spiral wave and HC modes temporally coexist in small systems but do not switch between each other in large systems.
Conversely, when multiple four-state cycles exist in six-state and cubic networks,
one state remains dominant for the entire range of flip energies, whereas the other states occasionally form domains at intermediate flip energies.
Therefore, the number of spatially coexisting states can be controlled using flip networks and energies.
\end{abstract} 

\maketitle

\section{Introduction}

Various types of pattern formations have been observed in nonequilibrium systems~\cite{nico77,hake04,mikh94,murr03,kura84,aceb05,kond21,nogu24c,beta17,bail22}.
These are called self-organizing systems~\cite{nico77,hake04}, dissipative structures~\cite{nico77},  synergetics~\cite{hake04,mikh94}, 
and nonreciprocal systems~\cite{you20,fruc21,rana24,guis24}, featuring different aspects.
The interactions between single oscillators or cyclic loops of states can generate spatiotemporal patterns and global oscillations.
Lattice models, such as the lattice Lotka--Volterra models~\cite{szol14,szab02,reic07,szcz13,kels15,dobr18,szab04,szab08,roma12,rulq14,baze19,zhon22,yang23,szol23} and active Potts models~\cite{nogu24a,nogu24b,nogu25,nogu25a,nogu25b,nogu25c}, have been widely used to study spatiotemporal patterns, such as traveling waves.
Nonequilibrium phase transitions have been studied using locally and globally coupled models.
The transition between the disordered phase and global oscillation has been reported to be of second order in the $XY$ universality class~\cite{risl04,wood06,avni25}.
In the six-state active Potts model, the values of the scaling exponents for the transition between the three-state wave and the six-state mixed modes
changed from their equilibrium values,
but no change was detected for the transition between the cyclic mode of the homogeneous phases and the three-state mixed mode~\cite{nogu25b}.
Dissipation~\cite{herp18,meib24} and mutual information~\cite{ptas25} were analyzed in active (driven) standard Potts models with global coupling. 

\begin{figure}[tbh]
\includegraphics[]{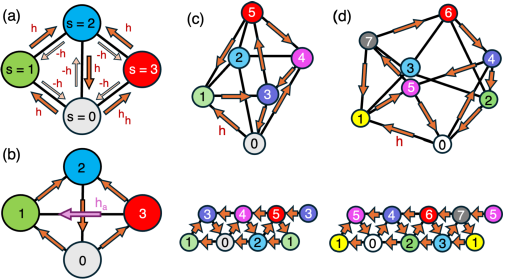}
\caption{
Flip networks comprising multiple three-state cycles in active Potts models.
(a) Four-state model with two three-state cycles.
The broad and narrow arrows represent the forward and backward flips with flip energies $h$ and $-h$, respectively.
(b) Four-state model with the tetrahedral network.
The flips between $s=1$ and $s=3$ are added in the model shown in (a) with 
flip energies $h_{\mathrm {a}}$ and $-h_{\mathrm {a}}$, respectively.
(c) Six-state model with the octahedral network.
(d) Eight-state model with the square-antiprism network.
The arrows for backward flips are omitted in (b)--(d).
The unfolded graphs are also shown in the bottom panels of (c) and (d). 
}
\label{fig:cart0}
\end{figure}

This study aims to clarify the effects of competition between identical cyclic loops of states on spatiotemporal patterns
using Monte Carlo (MC) simulations.
Previous studies typically considered a single oscillator or a cyclic loop of states at each site.
Previously, we investigated the competition between the three- and four-state loops in a four-state active standard Potts model~\cite{nogu25}.
Continuous changes between the three- and four-state dynamic modes were obtained by varying the flip ratios.
In this study, we consider multiple identical loops of states. The minimum is a four-state network comprising two three-state cycles shown in Fig.~\ref{fig:cart0}(a).
We examine three types of three-state cycle networks and two types of four-state cycle networks.
The dynamics differ from those of the single-cycle models.
In biological and social systems, complicated networks are observed (e.g., chemical reactions and gene expressions in cells)~\cite{beta17,bail22,nova08,aren08,arti24}.
The presented networks are simplified models.

The models and methods are described in Sec.~\ref{sec:model}.
Simulation results for the three- and four-state-cycle networks  are presented and discussed in  Secs.~\ref{sec:3} and \ref{sec:4}, respectively.
The dynamics for the single three-state and four-state cycles are briefly reviewed in Secs.~\ref{sec:3s} and \ref{sec:4s}, respectively.
The results of the four-state models with two three-state cycles are described in  Sec.~\ref{sec:3w}.
A theoretical analysis at the absence of contact interactions is presented in  Sec.~\ref{sec:3t}.
The results of the six- and eight-state models with the octahedral and square-antiprism networks are described in  Secs.~\ref{sec:3o} and \ref{sec:3ap}, respectively.
The  results of six- and eight-state models with two four-state cycles and a cubic network are described  in  Secs.~\ref{sec:4w} and \ref{sec:4c}, respectively.
Finally, a summary is presented in Sec.~\ref{sec:sum}.

\section{Models and Methods}\label{sec:model}

Two-dimensional square lattices with a side length $L$ are employed under periodic boundary conditions.
The total number of sites is $N=L^2$.
In a $q$-state Potts model, each site possesses a state $s\in [0,q-1]$.
The nearest neighboring sites ($i$ and $j$) have contact energies $J_{s_is_j}$:~\cite{wu82,pott52}
\begin{equation}
\label{eq:hint}
H_{\mathrm{int}} = - \sum_{\langle ij\rangle} J_{s_is_j}.
\end{equation}
In equilibrium systems, each state may additionally possess self-energy $\varepsilon_s$,
and the ratio of the forward and backward flip rates for the flipping of a single site from $s$ to $s'$ is given by $\exp(-\Delta H_{s_is'_i})$, where $\Delta H_{s_is'_i} = \Delta H_{\mathrm{int}} - h_{s,s'}$ is the energy difference between the two states, 
$h_{s,s'} = \varepsilon_s - \varepsilon_s'$ is the flip energy,
and the thermal energy $k_{\mathrm{B}}T$ is normalized to unity.
At equilibrium, the sum of flip energies along a flip cycle vanishes.
This model is extended to a nonequilibrium framework with nonzero cyclic sums, while maintaining $h_{ss'}=-h_{s's}$~\cite{nogu24a}.
Therefore, the detailed balance can be locally satisfied for flips between neighboring states, but not globally. 
For three-state cycles, this corresponds to the rock--paper--scissors relationship.
These types of dynamics can be realized in reactions on a catalytic surface~\cite{ertl08,bar94,goro94,barr20,zein22} and molecular transport across membranes~\cite{tabe03,miel20,holl21,nogu23}.
Active Potts models with global coupling were studied by Esposito et al~\cite{herp18,meib24,ptas25}. They called them driven Potts~\cite{herp18,meib24} and nonequilibrium Potts models~\cite{ptas25}.
The four-state active vector-Potts model is also called the nonreciprocal Ising~\cite{avni25} and nonreciprocal Ashkin--Teller model~\cite{nogu25b}.

\begin{figure}[tbh]
\includegraphics[]{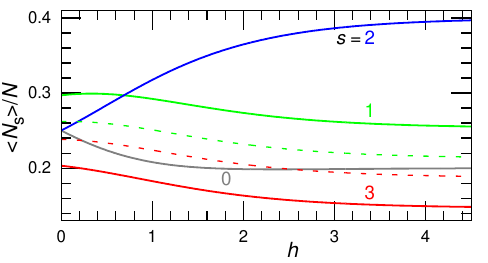}
\caption{
Flip energy $h$ dependence of the state densities $N_s/N$ for the tetrahedron
network [Fig.~\ref{fig:cart0}(b)].
The gray, green, blue, and red lines represent the data at $s=0$, $1$, $2$, and $3$, respectively.
The solid and dashed lines for $s=1$ and $3$ represent the data at $h_{\mathrm{a}}=1$ and $0.2$, respectively.
}
\label{fig:fix}
\end{figure}

In this study, the contact energy of the standard Potts model, $J_{ss'}=J\delta_{ss'}$, is used.
The dynamics under different types of contact interactions were reported in Refs.~\onlinecite{nogu25b,nogu25c}.
We set $J=2$ to induce the phase separation of different states, since
it is greater than the transition point $J_{\mathrm{c}}= \ln(1+ \sqrt{q})$ between the disordered and ordered phases at thermal equilibrium for $q=3$--$8$~\cite{wu82,baxt73}.
To construct identical cyclic loops, the flip energy is uniform as $h_{k,k'}=h$ in most cases.
The exception is the flip energy  $h_{\mathrm {a}}$ for the additional diagonal flips in the tetrahedral network, as shown in Fig.~\ref{fig:cart0}(b).

In the MC simulations,
a site is randomly selected, and
its flip to a neighboring state is performed using the Metropolis MC algorithm  with the acceptance ratio calculated as $p_{s_is'_i}=\min[1, \exp(-\Delta H_{s_is'_i})]$.
This flipping is attempted $N$ times per MC step (time unit).
We previously verified that the choice between the Metropolis and Glauber schemes for the MC update causes only a minor difference~\cite{nogu24a,nogu25c}.
The statistical errors are calculated from three or more independent runs.

\section{Competition among Three-state Cycles}\label{sec:3}

\subsection{Single Three-state Cycle}\label{sec:3s}

Before discussing the competition among multiple three-state cycles,
we briefly describe the dynamics induced by the single three-state cycle in the three-state active Potts model under cyclic symmetry~\cite{nogu24a,nogu24b}.
Under thermal equilibrium ($h=0$), one of the states  dominantly occupies the entire lattice,
and the three phases $s=0,1,2$ are equivalent.
At low $h$ ($h>0$), these three homogeneous phases change cyclically as $s=0\to 1\to 2\to 0$
through nucleation and growth.
This mode is called homogeneous cycling (HC).
At high $h$, the three states coexist spatially, and these domains form spiral waves.
The spiral centers are the contact points of the three domains. 
In this study, we call this mode W3 (W$n$ for waves comprising $n$ states).
At intermediate $h$, these two modes temporally coexist and stochastically switch
in small systems but do not switch in the available simulation periods in large systems
owing to the exponential increase in the lifetime of the W3 mode with the system size.

\subsection{Four-state Potts Model with Two Three-state Cycles}\label{sec:3w}

\subsubsection{Homogeneous Mixed State in the Absence of Interactions}\label{sec:3t}

We consider two types of flip networks, as shown in Fig.~\ref{fig:cart0}(a) and (b).
The first network (two triangles) has two competitive cycles ($s=0\to 1\to 2 \to 0$ and $s=0\to 3\to 2 \to 0$).
The two cycles are identical, and all flip energies are constant ($h$).
The second network has an additional flip between $s=1$ and $3$ 
with a flip energy of $h_{\mathrm{a}}$ [Fig.~\ref{fig:cart0}(b)].
These four states can be mapped onto the vertices of the tetrahedron.
When the edge lengths are proportional to the flip energies,
it is a regular tetrahedron at $h_{\mathrm{a}}=h$.

Before simulating the spatiotemporal patterns,
we investigated the flip dynamics of a homogeneous mixed state,
in which neighboring sites have no interactions ($J=0$).
Each site flips independently; hence,
the density $\rho_i= N_i/N$ of the $s=i$ state is governed by the equation,
\begin{equation}\label{eq:stt}
\frac{d\rho_i}{dt} = \sum_{j\ne i} w_{ji}\rho_j - w_{ij}\rho_i,
\end{equation}
where $w_{ij}$ is the flip rate from $s=i$  to $s=j$
and $\sum_i \rho_i = 1$.
The steady-state densities are given by
\begin{eqnarray}\label{eq:st20}
\frac{\rho_2}{\rho_0} &=& \frac{2+\exp(-h)+\exp(-2h)}{1+\exp(-h)+2\exp(-2h)}, \\ \label{eq:st130}
\frac{\rho_1+\rho_3}{2\rho_0} &=& \frac{[1+\exp(-h)]^2}{1+\exp(-h)+2\exp(-2h)}, \\ \label{eq:st13}
\frac{\rho_1}{\rho_3} &=& \frac{3+\exp(-h)}{1+\exp(-h)+2\exp(-h_{\mathrm{a}})},
\end{eqnarray}
using $d\rho_i/dt=0$ and the Metropolis rate $w_{ij}=\min[1,\exp(h_{ij})]$.
Interestingly, $\rho_0$, $\rho_2$, and $\rho_1+\rho_3$ are independent of $h_{\mathrm{a}}$ in the tetrahedral
network [Fig.~\ref{fig:cart0}(b)], and their values are given by Eqs.~(\ref{eq:st20}) and (\ref{eq:st130}) for both networks (see Fig.~\ref{fig:fix}). At $h\to \infty$ (i.e., when there are no backward flips), 
$\rho_2/\rho_0=2$ and $(\rho_1+\rho_3)/2\rho_0=1$.
The ratio $\rho_1/\rho_3$ increases with $h_{\mathrm{a}}$ and saturates at the maximum value $\rho_1/\rho_3=3$ as $h_{\mathrm{a}}\to \infty$ and $h\to \infty$.

\begin{figure}[tbh]
\includegraphics[]{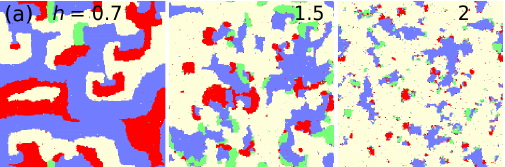}
\includegraphics[]{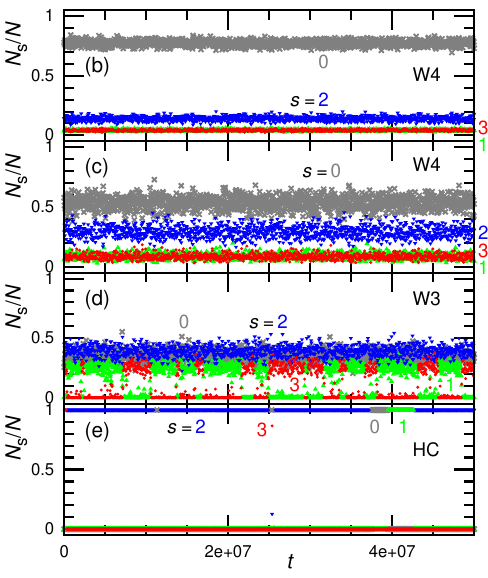}
\caption{
Active Potts model with two three-state cycles without diagonal flips [Fig.~\ref{fig:cart0}(a)] at $L=256$.
(a) Snapshots at $h=0.7$, $1.5$, and $2$ (from left to right).
The light yellow, green, blue, and red sites (light to dark in grayscale)
represent $s=0$, $1$, $2$, and $3$, respectively.
(b)--(e) Time development of the number of each state at (b) $h=2$, (c) $h=1.5$, and (d),(e) $h=0.7$.
Two three-state spiral waves ($s=0\to 1 \to 2 \to 0$, and $s=3$ instead of $s=1$)  spatially coexist (W4) at $h\gtrsim 0.9$, and the $s=0$ state becomes dominant at high $h$.
The three-state spiral waves (W3) and homogeneous cycling (HC) are obtained at $h=0.7$, depending on the initial state.
In W3, the two types of waves temporally coexist as shown in (d). 
}
\label{fig:q4t}
\end{figure}

\begin{figure}[tbh]
\includegraphics[]{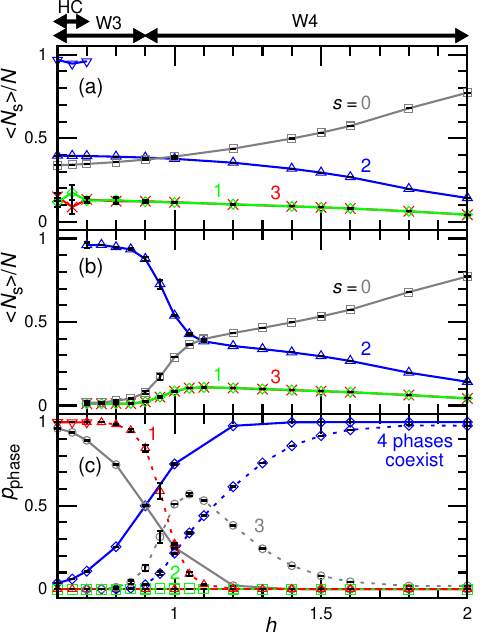}
\caption{
Dependence on the flip energy $h$ in the active Potts model with two three-state cycles without diagonal flips [Fig.~\ref{fig:cart0}(a)].
(a)--(b) Mean number density $\langle N_s\rangle/N$ at (a) $L=256$ and (b) $L=128$.
The blue down-pointing triangles in (a) represent $s=2$ in the HC mode.
(c) Time fractions $p_{\mathrm{phase}}$ of phases at $L=256$ (solid lines) and $L=128$ (dashed lines).
The red triangles represent the ratio of the single-state dominant phase in the HC mode at $L=256$,
whereas the ratios of multiple-state coexisting phases in the HC mode are omitted for clarity.
The bidirectional arrows at the top of (a) represent the ranges of three modes at $L=256$.
}
\label{fig:q4p}
\end{figure}

\begin{figure}[tbh]
\includegraphics[]{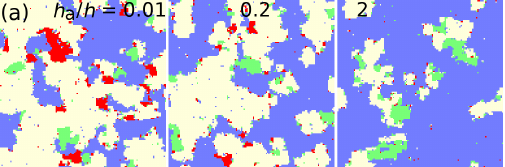}
\includegraphics[]{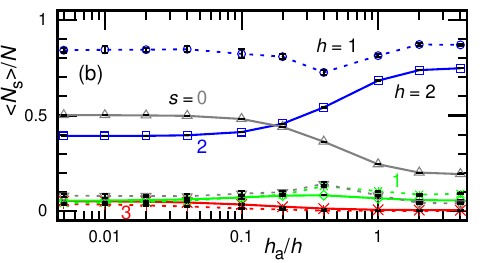}
\caption{
Active Potts model with two three-state cycles and diagonal flips [Fig.~\ref{fig:cart0}(b)] at $L=128$.
(a) Snapshots for $h_{\mathrm {a}}/h=0.01$, $0.2$, and $2$ at $h=2$ (from left to right).
(b) Mean number density $\langle N_s\rangle/N$ as a function of $h_{\mathrm{a}}/h$.
The solid and dashed lines represent the data at $h=2$ and $1$, respectively.
}
\label{fig:q4ha}
\end{figure}

\subsubsection{Dynamics in the Absence of Diagonal Flips}\label{sec:3wa}

We consider herein the dynamics under contact interactions ($J=2$) for two three-state cycles [Fig.~\ref{fig:cart0}(a)].
At low $h$, cyclic changes in homogeneous phases occur via two routes ($s=0\to 1\to 2 \to 0$ and $s=0\to 3\to 2 \to 0$) i.e., the HC mode emerges [the numbers of states $N_s/N\approx 1$ are switched in Fig.~\ref{fig:q4t}(e)].
At high $h$, the domains of all four states spatially coexist and two types of spiral waves ($s=0\to 1\to 2 \to 0$ and $s=0\to 3\to 2 \to 0$) 
are steadily formed (W4) [see Fig.~\ref{fig:q4t}(b),(c) and the right two snapshots in Fig.~\ref{fig:q4t}(a)].
At intermediate $h$, one type of spiral wave is dominantly formed (W3), and the switch to the other type occurs stochastically [the positions of the red diamonds and green triangles are switched several times in Fig.~\ref{fig:q4t}(d)].
In the spiral waves shown in the left snapshot of Fig.~\ref{fig:q4t}(a) and Movie S1, small domains ($s=1$) often appear due to nucleation, but most of them eventually  disappear owing to fluctuations.
With increasing $h$, the HC and W3 modes begin to coexist temporally, and the ratio of the W3 mode gradually increases at $L=128$.
Conversely, the switching between two modes does not occur in available simulation periods ($\sim 10^8$) at $L=256$;
the HC and W3 modes are obtained when the uniform and random distributions of states are respectively used as initial states
for $0.3 \lesssim h \lesssim 0.7$.

To quantify these dynamic behaviors,
we calculate the mean densities of states $\langle N_s\rangle/N$ and the time fractions $p_{\mathrm{phase}}$ of 
single phases and multiple-phase coexistence (see Fig.~\ref{fig:q4p}).
The lattice is considered to be covered by a single phase at $N_s/N>0.98$ for $s \in [0,q-1]$,
and $n$ phases  coexist spatially when $n$ states satisfy $N_s/N>0.02$.
The ratios of the three- and four-state coexistence phases continuously change with increasing $h$ for both $N=128$ and $256$, as shown in Fig.~\ref{fig:q4p}(c).
In the temporal coexistence range of the HC and W3 modes at $N=128$,
the ratios of the single-state dominant and three-state coexistence phases gradually change, reflecting the time fractions of the two modes.
Hence, the HC, W3, and W4 modes are considered dominant
when the single-state dominant and three- and four-state coexistence phases have the highest fraction $p_{\mathrm{phase}}$, respectively
 (see the bidirectional arrows at the top of Fig.~\ref{fig:q4p} for $N=256$).

Interestingly, the $h$ dependence of the mean densities of states is different from that in homogeneous conditions [compare Figs.~\ref{fig:q4p}(a),(b) and \ref{fig:fix}].
The density of the $s=0$ state ($\langle N_0\rangle/N$) increases with increasing $h$ in Fig.~\ref{fig:q4p}(a),(b),
whereas the density of the $s=2$ state increases in Fig.~\ref{fig:fix}.
When the receiving and sending forward flips are counted as $+1$ and $-1$ flips for each state,
the numbers of flips for the $s=2$ and $s=0$ states are $+1$ and $-1$ flips, respectively.
Hence, $\langle N_2\rangle/N$ increases with $h$ in the absence of the contact interactions,
and $\langle N_0\rangle/N$ decreases.
In the HC mode, nucleation can occur twice in the ($s=0$)-dominant phase (compared with the $s=2$ phase),
such that the lattice is mostly covered by the $s=2$ phase.
However, in the wave modes, the flips mainly occur at the domain boundaries;
 hence, the flips from the $s=0$ state to the $s=1$ and $s=3$ states become slower than those from $s=2$ to $s=0$,
resulting in an increase in $\langle N_0\rangle/N$.

\begin{figure}[tbh]
\includegraphics[]{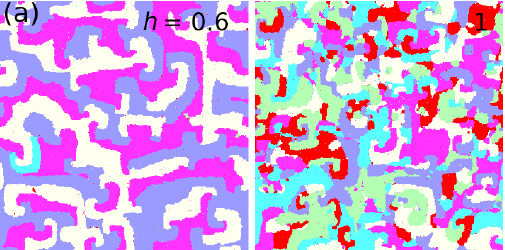}
\includegraphics[]{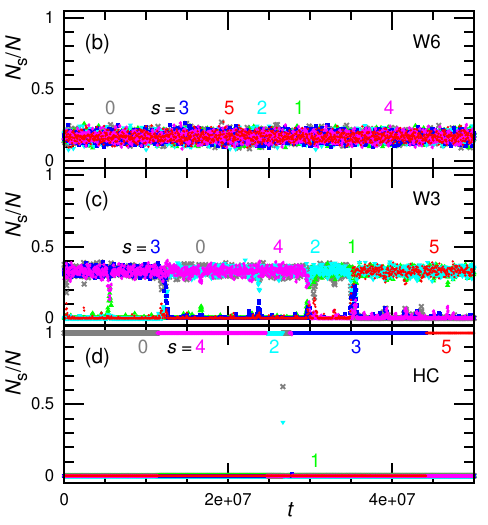}
\caption{
Active Potts model with the octahedral network [Fig.~\ref{fig:cart0}(c)] at $L=512$.
(a) Snapshots at $h=0.6$ (left) and $h=1$ (right).
The light yellow, light green, cyan, blue, magenta, and red sites (light to dark in grayscale) 
represent $s=0$, $1$, $2$, $3$, $4$, and $5$, respectively.
(b)--(d) Time development of the number of each state at (b) $h=1$ and (c),(d) $h=0.6$.
All types of three-state spiral waves  spatially coexist (W6) at $h= 1$.
The three-state spiral waves (W3) and homogeneous cycling (HC) are obtained at $h=0.6$, depending on the initial state,
[(c) and (d), respectively].
In W3, the eight types of waves temporally coexist. 
}
\label{fig:octat}
\end{figure}

\subsubsection{Dynamics in the Presence of Diagonal Flips}\label{sec:3wb}

When diagonal flips between the $s=1$ and $3$ states are added [Fig.~\ref{fig:cart0}(b)],
the number density of the $s=3$ state decreases with increasing $h_{\mathrm{a}}$, as in the case in which there are no contact interactions [compare Figs.~\ref{fig:q4ha}(b) and \ref{fig:fix}].
However, the densities of the $s=2$ and $s=0$ states also change in the presence of diagonal flips,
unlike the values in their absence.
For $h=2$, the $s=2$ density monotonically increases with increasing $h_{\mathrm{a}}$,
whereas a minimum appears at $h_{\mathrm{a}}/h \simeq 0.4$ for $h=1$ owing to a decrease in the time fraction of the HC mode.
The flips from the $s=3$ to $s=1$ states yield small clusters of $s=1$ and $3$ on the domain boundary of $s=0$,
so that the flips from the $s=0$ to $s=1$ or $3$ are enhanced,  resulting in a decrease in $\langle N_0\rangle/N$ (see Fig.~\ref{fig:q4ha}).

\begin{figure}[tbh]
\includegraphics[]{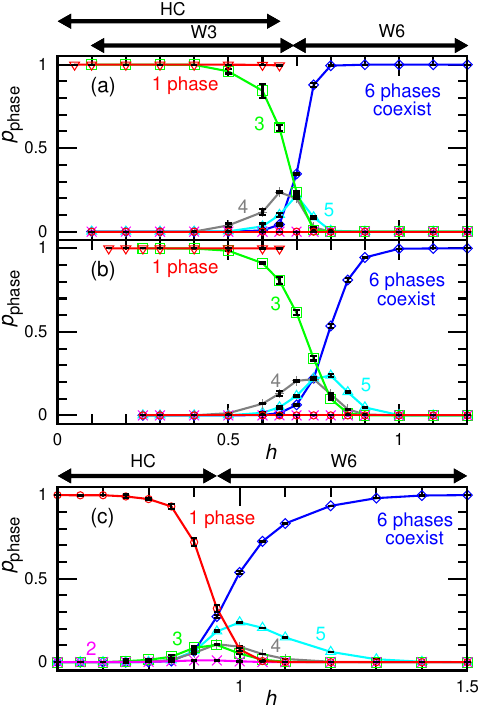}
\caption{
Phase ratios $p_{\mathrm{phase}}$ as functions of $h$ at (a) $L=512$, (b) $L=256$, and (c) $L=128$
in the active Potts model with the octahedral network [Fig.~\ref{fig:cart0}(c)].
The red down-pointing triangles in (a) and (b) 
represent the ratio of the single-state dominant phase in the HC mode,
whereas the ratios of multiple-state coexisting phases in the HC mode are omitted for clarity.
The bidirectional arrows at the tops of (a) and (c) represent the modes at $L=512$ and $128$, respectively.
}
\label{fig:octap}
\end{figure}

\begin{figure}[tbh]
\includegraphics[]{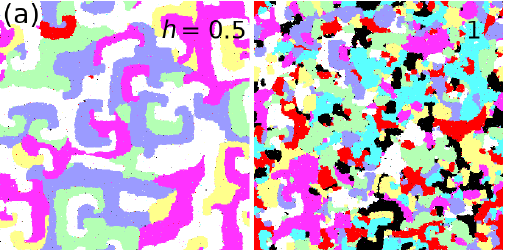}
\includegraphics[]{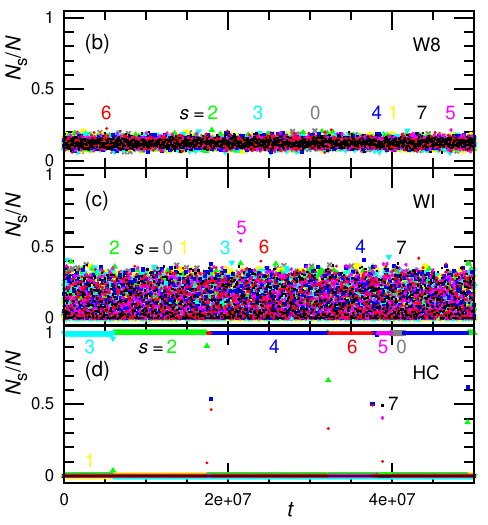}
\caption{
Active Potts model with the square-antiprism network [Fig.~\ref{fig:cart0}(d)] at $L=512$.
(a) Snapshots at $h=0.5$ (left) and $h=1$ (right).
The white, yellow, light green, cyan, blue, magenta, red, and black sites (light to dark in grayscale)
represent $s=0$, $1$, $2$, $3$, $4$, $5$, $6$, and $7$, respectively.
(b)--(d) Time development of the number of each state at (b) $h=1$, (c) $h=0.5$, and (d) $h=0.6$.
(b) All types of spiral waves comprising the eight states  spatially coexist (W8) at $h= 1$.
(c) Several (but not all) spiral waves spatially coexist (WI) at $h= 0.5$.
(d) The HC mode is obtained at $h \lesssim 0.6$, depending on the initial state.
}
\label{fig:ap4t}
\end{figure}

\begin{figure}[tbh]
\includegraphics[]{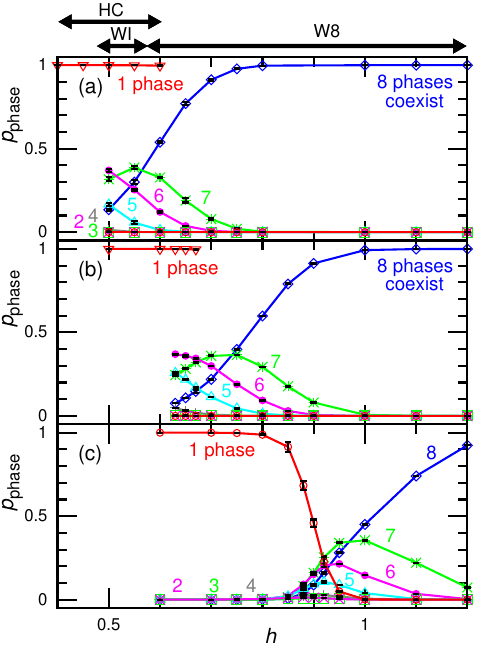}
\caption{
Phase ratios $p_{\mathrm{phase}}$ as functions of $h$ at (a) $L=512$, (b) $L=256$, and (c) $L=128$
in the active Potts model with the square-antiprism network [Fig.~\ref{fig:cart0}(d)].
The red down-pointing triangles in (a) and (b) 
represent the ratio of the single-state dominant phase in the HC mode,
whereas the ratios of multiple-state coexisting phases in the HC mode are omitted for clarity.
The bidirectional arrows at the top of (a) represent the modes at $L=512$.
}
\label{fig:ap4p}
\end{figure}

\subsection{Six-state Potts Model with Octahedral Network}\label{sec:3o}

We consider herein the octahedral network shown in Fig.~\ref{fig:cart0}(c).
All six states are identical, and all eight faces represent three-state cycles.
Similar to the four-state models described in Sec.~\ref{sec:3wa}, 
the HC, W3, and W6 modes emerge from low to high $h$ (see Fig.~\ref{fig:octat} and Movie S2).
In the HC and W3 modes, the composing states stochastically change into neighboring states and cycles, respectively.

In large systems ($L=256$ and $512$), hysteresis is obtained between the HC and W3 modes,
whereas the transition between the W3 and W6 modes is continuous [see Fig.~\ref{fig:octap}(a) and (b)].
A larger system has a broader hysteresis range of $h$.
In small systems ($L=128$), the HC mode changes directly to the W6 mode via the temporal coexistence of the two modes [see Fig.~\ref{fig:octap}(c)].

\subsection{Eight-state Potts Model with Square-Antiprism Network}\label{sec:3ap}

We consider herein a larger network of identical states.
The octahedron is the triangular antiprism,
and the next smallest antiprism is the square antiprism shown in Fig.~\ref{fig:cart0}(d).
Two triangles are added horizontally in the unfolded graph [compare the bottom panels in Fig.~\ref{fig:cart0}(c) and (d)].
In the square antiprism,
all eight states are identical, 
eight faces represent three-state cycles, and two faces represent four-state cycles.

Similar to the octahedral network model described in Sec.~\ref{sec:3o},
the HC and W8 modes occur at low and high $h$, respectively,
and the HC mode exhibits hysteresis in large systems ($L=256$ and $512$),
whereas the transition of the HC mode occurs via temporal coexistence in small systems ($L=128$),
as shown in Figs.~\ref{fig:ap4t} and \ref{fig:ap4p}.
The difference is that more than three states (but not all) exist in the spiral waves at low $h$ in large systems 
[see the left panel of Fig.~\ref{fig:ap4t}(a) and Movie S3].
The state densities $N_s/N$ are widely distributed, including $N_s/N\approx 0$ [see Fig.~\ref{fig:ap4t}(c)].
This is due to the existence of several separated cycles;
for example, the $s=0 \to 1\to 5$ cycle is separated from the cycles of $s=4 \to 2\to 6$, $s=2 \to 6\to 3$, and $s=6 \to 3\to 7$ [see Fig.~\ref{fig:cart0}(d)].
In contrast, the diagonal cycles are only separated in the octahedral network (e.g., $s=0\to 1\to 3$ and $s=2\to 5\to 4$) [see Fig.~\ref{fig:cart0}(c)].
The domain boundaries of non-neighboring states do not move ballistically (see Movies S3 and S4),
so that the spiral waves of separated cycles can coexist for a longer duration than those of neighboring cycles.

In all three networks,
spiral waves of a single three-state cycle or multiple cycles emerge.
Therefore, the spiral wave is a robust mode for the three-state cycles.
The number of cycles and states can vary according to the network and flip energy choices.
The competition among the cycles reduce the number of spatially coexisting states
at intermediate flip energies.

\begin{figure}[tbh]
\includegraphics[]{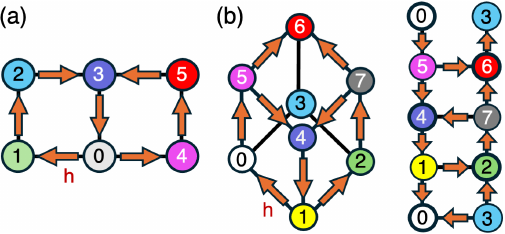}
\caption{
Flip networks comprising multiple four-state cycles in active Potts models.
The arrows for backward flips are omitted.
(a) Six-state model with two four-state cycles.
(b) Eight-state model with the cubic network.
The unfolded graph is shown in the right panel.
}
\label{fig:cart1}
\end{figure}

\begin{figure}[tbh]
\includegraphics[]{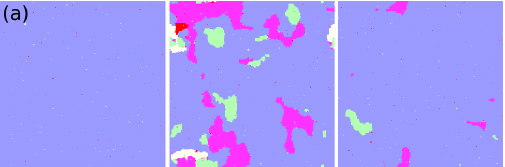}
\includegraphics[]{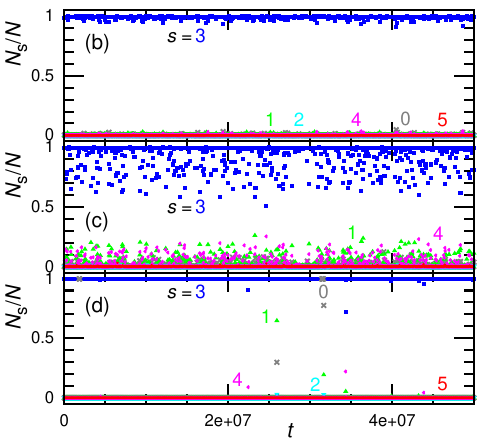}
\caption{
Active Potts model with two four-state cycles [Fig.~\ref{fig:cart1}(a)] at $L=256$.
(a) Sequential snapshots with a time interval of $\Delta t = 5000$ at $h=1.1$.
(b)--(d) Time development of the number of each state at (b) $h=1.4$, (c) $h=1.1$, and (d) $h=0.7$.
}
\label{fig:q6rt}
\end{figure}

\begin{figure}[tbh]
\includegraphics[]{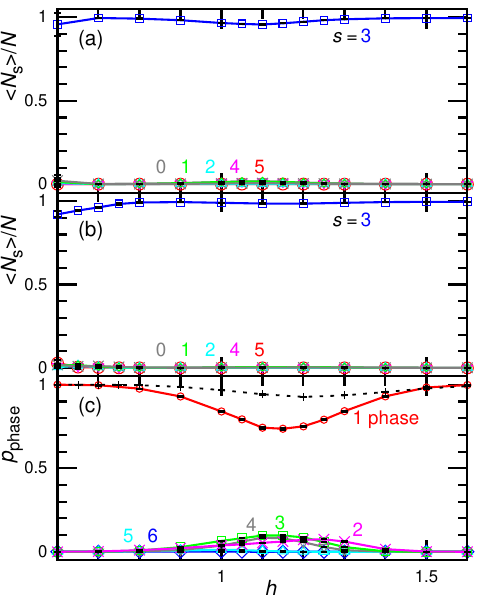}
\caption{
Dependence on the flip energy $h$ in the active Potts model with two four-state cycles [Fig.~\ref{fig:cart1}(a)].
(a)--(b) Mean numbers of states $\langle N_s\rangle/N$ at (a) $L=256$ and (b) $L=128$.
(c) Phase ratios $p_{\mathrm{phase}}$.
The solid lines represent the data at $L=256$.
The dashed line represents the ratio of the single-state dominant phase  at $L=128$,
whereas the ratios of multiple-state coexisting phases at $L=128$ are omitted for clarity.
}
\label{fig:q6rp}
\end{figure}

\section{Competition among Four-state Cycles}\label{sec:4}

\subsection{Single Four-state Cycle}\label{sec:4s}

Before discussing the competition among multiple four-state cycles,
we briefly describe the dynamics induced by the single four-state cycle in the four-state active standard Potts model under cyclic symmetry~\cite{nogu25}.
At low $h$ ($h>0$), the four homogeneous phases change cyclically as $s=0\to 1\to 2\to 3 \to 0$
through nucleation and growth (HC mode), as in the three-state model.
At high $h$, the four states spatially coexist, but the waves  do not have spiral centers (W4 mode)
under the interaction of the standard Potts model~\cite{nogu25,nogu25b}.
The domain boundaries between neighboring states propagate ballistically, but the boundaries between non-neighboring  states move
 slowly (driven by diffusion).
Hence, the existence of non-flip diagonal states alters the dynamics.
The transition between the HC and W4 modes occurs via temporal coexistence,
even at $L=512$.
At intermediate flip energies,
the domains of the diagonal states often remain for long periods,
 slowly shrinking to reduce the interfacial energy.

\begin{figure}[tbh]
\includegraphics[]{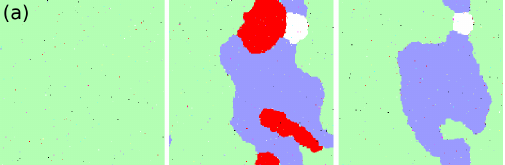}
\includegraphics[]{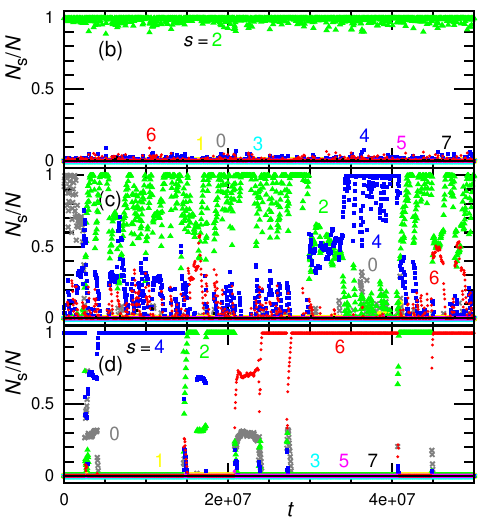}
\caption{
Active Potts model with the cubic network [Fig.~\ref{fig:cart1}(b)] at $L=256$.
(a) Sequential snapshots with a time interval of $\Delta t = 40000$ at $h=0.9$.
(b)--(d) Time development of the number of each state at (b) $h=1.1$, (c) $h=0.9$, and (d) $h=0.7$.
}
\label{fig:q8cubt}
\end{figure}

\begin{figure}[tbh]
\includegraphics[]{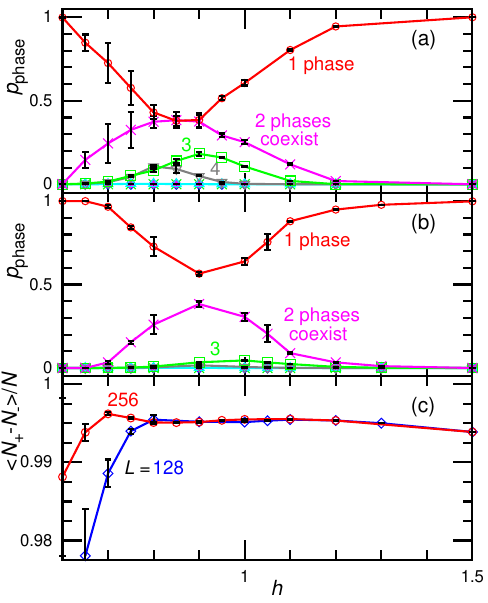}
\caption{
Dependence on the flip energy $h$ in the active Potts model with the cubic network [Fig.~\ref{fig:cart1}(b)]. 
(a)--(b) Phase ratios $p_{\mathrm{phase}}$ at (a) $L=256$ and (b) $L=128$.
(c) Mean number differences $\langle N_+-N_-\rangle/N$ between even- and odd-numbered states at $L=256$ and $128$, where $N_+=N_0+N_2+N_4+N_6$ and  $N_-=N_1+N_3+N_5+N_7$.
}
\label{fig:q8cubp}
\end{figure}

\subsection{Six-state Potts Model with Two Four-state Cycles}\label{sec:4w}

First, we consider the competition between two four-state cycles ($s=0\to 1\to 2\to 3\to 0$ and $s=0\to 4\to 5\to 3\to 0$) 
in the six-state model shown in Fig.~\ref{fig:cart1}(a).
The $s=3$ state always covers the lattice for most of the time,
unlike in the three-state cycles [see Figs.~\ref{fig:q6rt} and \ref{fig:q6rp}].
Although the HC mode appears at low $h$, the ($s=3$)-dominant phase has the longest lifetime.
At intermediate $h$,
the nucleation of the $s=0$ state results in the formation of $s=1$ and $4$ domains in the $s=3$ phase. These domains slowly shrink as shown in Fig.~\ref{fig:q6rt} and Movie S5.
These domains become smaller with increasing $h$ [compare Fig.~\ref{fig:q6rt}(b) and (c)].
Thus, traveling waves do not form steadily under any conditions in this network.

\subsection{Eight-state Potts Model with Cubic Network}\label{sec:4c}

We consider herein an eight-state model with a cubic network, as shown in Fig.~\ref{fig:cart1}(b).
This network has two types of states:
the four even-numbered states ($s=0$, $2$, $4$, and $6$)
receive two forward flips and one backward flip,
while the odd-numbered states ($s=1$, $3$, $5$, and $7$)
receive one forward flip and two backward flips.

The even-numbered states dominantly covered the lattice,
as the $s=3$ state in the case of the six-state model described in Sec.~\ref{sec:4w} (see Figs.~\ref{fig:q8cubt} and \ref{fig:q8cubp}).
The HC mode appears at low $h$ [see Fig.~\ref{fig:q8cubt}(d)],
and diagonal domains are formed at intermediate and high $h$ [see Fig.~\ref{fig:q8cubt}(a)--(c) and Movie S6].
In contrast to the six-state model, these domains are larger and have longer lifetimes.
This is due to the existence of more diagonal states (e.g., $s=2$, $4$, $6$, and $7$ for $s=0$),
and three of them exhibit the same stability ($s=2$, $4$, and $6$ for $s=0$).
This domain formation also occurs in the HC mode [see Fig.~\ref{fig:q8cubt}(d)].

In both networks,
the competition among the four-state cycles
suppresses the traveling waves. 
The single-state dominant phases are stabilized, and the domains of the nonflip (diagonal) states
 often coexist. 
Therefore, the existence of nonflip states in the flip cycles alters the dynamics.

\section{Summary}\label{sec:sum}

We studied the spatiotemporal patterns in active Potts models with multiple flip cycles.
For the competition among three-state cycles,
three networks were examined: a four-state network of two cycles (with and without diagonal flips),
a six-state octahedral network, and an eight-state square-antiprism network.
At low flip energy $h$, the changes in single-state dominant phases
 along the network bonds occur for all networks (HC mode).
At high $h$, all states spatially coexist as spiral waves in all cycles.
At intermediate $h$, spiral waves of single three-state cycles emerge in the cases of the first two networks,
whereas spiral waves of multiple (but not all) cycles emerge in the case of the antiprism network.
The transition from the HC and wave modes occurs via the temporal coexistence of the two modes in small systems.
However, the dynamics are trapped in either mode via hysteresis in large systems.

Regarding the competition among the four-state cycles,
two networks were examined: a six-state network of two cycles and an eight-state cubic network.
Single-state dominant phases occupy the systems most of the time for all simulation conditions,
so that the steady traveling waves are suppressed by competition.
At intermediate $h$,  diagonal states often form  domains but subsequently shrink slowly.
This is due to the absence of direct flips between the diagonal states in the four-state cycles.

Let us discuss how the flip-cycle competition changes spatiotemporal patterns.
In the competitive three-state cycles, 
the single-cycle spiral waves of three states are obtained at small networks (the four- and six-state networks)
with intermediate flip energies.
Emerged waves suppress the growth of neighboring states so that
spiral waves comprising other states do not spatially coexist.
However, cycles separated in the flip network can coexist at large networks,
and waves of different cycles can simultaneously appear.
The number of cycles in waves increases in a larger network.
Conversely, the competition among multiple four-state cycles suppresses all waves, leading to single-state dominant phases.
Therefore, the competition can be used to reduce the wave types or erase the waves using three- or four-state cycles, respectively.

In this study, we employed the contact interactions for the standard Potts model.
For a single cycle of four or more states, waves do not exhibit spiral shapes in the standard Potts model.
However, spiral waves are formed when the contacts of nonflip states are repulsive,
and factorized-symmetry modes also emerge for cyclic loops of factorizable numbers of states (e.g., $6=3\times 2$)~\cite{nogu25b}.
Therefore, the competition among spiral waves and other forms of dynamics can be explored when different contact interactions are used in active Potts models with multiple cycles of four or more states.

In this study, we used a constant flip energy for all flips (except for the diagonal ones for two three-state cycles).
By adjusting the flip and contact energies in asymmetric networks, spiral waves of specific cycles can be generated selectively.
Hence, multiple cycles and states in wave modes can be controlled by choosing the networks and conditions.

\begin{acknowledgments}
This work was supported by JSPS KAKENHI Grant Number JP24K06973. 
\end{acknowledgments}


\begin{thebibliography}{55}%
\makeatletter
\providecommand \@ifxundefined [1]{%
 \@ifx{#1\undefined}
}%
\providecommand \@ifnum [1]{%
 \ifnum #1\expandafter \@firstoftwo
 \else \expandafter \@secondoftwo
 \fi
}%
\providecommand \@ifx [1]{%
 \ifx #1\expandafter \@firstoftwo
 \else \expandafter \@secondoftwo
 \fi
}%
\providecommand \natexlab [1]{#1}%
\providecommand \enquote  [1]{``#1''}%
\providecommand \bibnamefont  [1]{#1}%
\providecommand \bibfnamefont [1]{#1}%
\providecommand \citenamefont [1]{#1}%
\providecommand \href@noop [0]{\@secondoftwo}%
\providecommand \href [0]{\begingroup \@sanitize@url \@href}%
\providecommand \@href[1]{\@@startlink{#1}\@@href}%
\providecommand \@@href[1]{\endgroup#1\@@endlink}%
\providecommand \@sanitize@url [0]{\catcode `\\12\catcode `\$12\catcode
  `\&12\catcode `\#12\catcode `\^12\catcode `\_12\catcode `\%12\relax}%
\providecommand \@@startlink[1]{}%
\providecommand \@@endlink[0]{}%
\providecommand \url  [0]{\begingroup\@sanitize@url \@url }%
\providecommand \@url [1]{\endgroup\@href {#1}{\urlprefix }}%
\providecommand \urlprefix  [0]{URL }%
\providecommand \Eprint [0]{\href }%
\providecommand \doibase [0]{http://dx.doi.org/}%
\providecommand \selectlanguage [0]{\@gobble}%
\providecommand \bibinfo  [0]{\@secondoftwo}%
\providecommand \bibfield  [0]{\@secondoftwo}%
\providecommand \translation [1]{[#1]}%
\providecommand \BibitemOpen [0]{}%
\providecommand \bibitemStop [0]{}%
\providecommand \bibitemNoStop [0]{.\EOS\space}%
\providecommand \EOS [0]{\spacefactor3000\relax}%
\providecommand \BibitemShut  [1]{\csname bibitem#1\endcsname}%
\let\auto@bib@innerbib\@empty
\bibitem [{\citenamefont {Nicolis}\ and\ \citenamefont
  {Prigogine}(1977)}]{nico77}%
  \BibitemOpen
  \bibfield  {author} {\bibinfo {author} {\bibfnamefont {G.}~\bibnamefont
  {Nicolis}}\ and\ \bibinfo {author} {\bibfnamefont {I.}~\bibnamefont
  {Prigogine}},\ }\href@noop {} {\emph {\bibinfo {title} {Self-organization in
  nonequilibrium systems : From dissipative structures to order through
  fluctuations}}}\ (\bibinfo  {publisher} {Wiley},\ \bibinfo {address} {New
  York},\ \bibinfo {year} {1977})\BibitemShut {NoStop}%
\bibitem [{\citenamefont {Haken}(2004)}]{hake04}%
  \BibitemOpen
  \bibfield  {author} {\bibinfo {author} {\bibfnamefont {H.}~\bibnamefont
  {Haken}},\ }\href@noop {} {\emph {\bibinfo {title} {Synergetics :
  Introduction and advanced topics}}}\ (\bibinfo  {publisher} {Springer},\
  \bibinfo {address} {Berlin},\ \bibinfo {year} {2004})\BibitemShut {NoStop}%
\bibitem [{\citenamefont {Mikhailov}(1994)}]{mikh94}%
  \BibitemOpen
  \bibfield  {author} {\bibinfo {author} {\bibfnamefont {A.}~\bibnamefont
  {Mikhailov}},\ }\href@noop {} {\emph {\bibinfo {title} {Foundations of
  synergetics I: Distributed active systems}}},\ \bibinfo {edition} {2nd}\ ed.\
  (\bibinfo  {publisher} {Springer},\ \bibinfo {address} {Berlin},\ \bibinfo
  {year} {1994})\BibitemShut {NoStop}%
\bibitem [{\citenamefont {Murray}(2003)}]{murr03}%
  \BibitemOpen
  \bibfield  {author} {\bibinfo {author} {\bibfnamefont {J.~D.}\ \bibnamefont
  {Murray}},\ }\href@noop {} {\emph {\bibinfo {title} {Mathematical biology II:
  Spatial models and biomedical applications}}},\ \bibinfo {edition} {3rd}\
  ed.\ (\bibinfo  {publisher} {Springer},\ \bibinfo {address} {New York},\
  \bibinfo {year} {2003})\BibitemShut {NoStop}%
\bibitem [{\citenamefont {Kuramoto}(1984)}]{kura84}%
  \BibitemOpen
  \bibfield  {author} {\bibinfo {author} {\bibfnamefont {Y.}~\bibnamefont
  {Kuramoto}},\ }\href@noop {} {\emph {\bibinfo {title} {Chemical oscillations,
  waves, and turbulence}}}\ (\bibinfo  {publisher} {Springer},\ \bibinfo
  {address} {Berlin},\ \bibinfo {year} {1984})\BibitemShut {NoStop}%
\bibitem [{\citenamefont {Acebr\'on}\ \emph {et~al.}(2005)\citenamefont
  {Acebr\'on}, \citenamefont {Bonilla}, \citenamefont {P\'erez~Vicente},
  \citenamefont {Ritort},\ and\ \citenamefont {Spigler}}]{aceb05}%
  \BibitemOpen
  \bibfield  {author} {\bibinfo {author} {\bibfnamefont {J.~A.}\ \bibnamefont
  {Acebr\'on}}, \bibinfo {author} {\bibfnamefont {L.~L.}\ \bibnamefont
  {Bonilla}}, \bibinfo {author} {\bibfnamefont {C.~J.}\ \bibnamefont
  {P\'erez~Vicente}}, \bibinfo {author} {\bibfnamefont {F.}~\bibnamefont
  {Ritort}}, \ and\ \bibinfo {author} {\bibfnamefont {R.}~\bibnamefont
  {Spigler}},\ }\href {\doibase 10.1103/RevModPhys.77.137} {\bibfield
  {journal} {\bibinfo  {journal} {Rev. Mod. Phys.}\ }\textbf {\bibinfo {volume}
  {77}},\ \bibinfo {pages} {137} (\bibinfo {year} {2005})}\BibitemShut
  {NoStop}%
\bibitem [{\citenamefont {Kondo}\ \emph {et~al.}(2021)\citenamefont {Kondo},
  \citenamefont {Watanabe},\ and\ \citenamefont {Miyazawa}}]{kond21}%
  \BibitemOpen
  \bibfield  {author} {\bibinfo {author} {\bibfnamefont {S.}~\bibnamefont
  {Kondo}}, \bibinfo {author} {\bibfnamefont {M.}~\bibnamefont {Watanabe}}, \
  and\ \bibinfo {author} {\bibfnamefont {S.}~\bibnamefont {Miyazawa}},\ }\href
  {\doibase 10.1098/rsta.2020.0274} {\bibfield  {journal} {\bibinfo  {journal}
  {Phil. Trans. R. Soc. A}\ }\textbf {\bibinfo {volume} {379}},\ \bibinfo
  {pages} {20200274} (\bibinfo {year} {2021})}\BibitemShut {NoStop}%
\bibitem [{\citenamefont {Noguchi}(2025{\natexlab{a}})}]{nogu24c}%
  \BibitemOpen
  \bibfield  {author} {\bibinfo {author} {\bibfnamefont {H.}~\bibnamefont
  {Noguchi}},\ }\href {\doibase 10.1002/syst.202400042} {\bibfield  {journal}
  {\bibinfo  {journal} {ChemSystemsChem}\ }\textbf {\bibinfo {volume} {7}},\
  \bibinfo {pages} {e202400042} (\bibinfo {year}
  {2025}{\natexlab{a}})}\BibitemShut {NoStop}%
\bibitem [{\citenamefont {Beta}\ and\ \citenamefont {Kruse}(2017)}]{beta17}%
  \BibitemOpen
  \bibfield  {author} {\bibinfo {author} {\bibfnamefont {C.}~\bibnamefont
  {Beta}}\ and\ \bibinfo {author} {\bibfnamefont {K.}~\bibnamefont {Kruse}},\
  }\href {\doibase 10.1146/annurev-conmatphys-031016-025210} {\bibfield
  {journal} {\bibinfo  {journal} {Annu. Rev. Condens. Matter Phys.}\ }\textbf
  {\bibinfo {volume} {8}},\ \bibinfo {pages} {239} (\bibinfo {year}
  {2017})}\BibitemShut {NoStop}%
\bibitem [{\citenamefont {Bailles}\ \emph {et~al.}(2022)\citenamefont
  {Bailles}, \citenamefont {Gehrels},\ and\ \citenamefont {Lecuit}}]{bail22}%
  \BibitemOpen
  \bibfield  {author} {\bibinfo {author} {\bibfnamefont {A.}~\bibnamefont
  {Bailles}}, \bibinfo {author} {\bibfnamefont {E.~W.}\ \bibnamefont
  {Gehrels}}, \ and\ \bibinfo {author} {\bibfnamefont {T.}~\bibnamefont
  {Lecuit}},\ }\href {\doibase 10.1146/annurev-cellbio-120420-095337}
  {\bibfield  {journal} {\bibinfo  {journal} {Annu. Rev. Cell Dev. Biol.}\
  }\textbf {\bibinfo {volume} {38}},\ \bibinfo {pages} {321} (\bibinfo {year}
  {2022})}\BibitemShut {NoStop}%
\bibitem [{\citenamefont {You}\ \emph {et~al.}(2020)\citenamefont {You},
  \citenamefont {Baskaran},\ and\ \citenamefont {Marchetti}}]{you20}%
  \BibitemOpen
  \bibfield  {author} {\bibinfo {author} {\bibfnamefont {Z.}~\bibnamefont
  {You}}, \bibinfo {author} {\bibfnamefont {A.}~\bibnamefont {Baskaran}}, \
  and\ \bibinfo {author} {\bibfnamefont {M.~C.}\ \bibnamefont {Marchetti}},\
  }\href {\doibase 10.1073/pnas.2010318117} {\bibfield  {journal} {\bibinfo
  {journal} {Proc.\ Natl.\ Acad.\ Sci.\ USA}\ }\textbf {\bibinfo {volume}
  {117}},\ \bibinfo {pages} {19767} (\bibinfo {year} {2020})}\BibitemShut
  {NoStop}%
\bibitem [{\citenamefont {Fruchart}\ \emph {et~al.}(2021)\citenamefont
  {Fruchart}, \citenamefont {Hanai}, \citenamefont {Littlewood},\ and\
  \citenamefont {Vitelli}}]{fruc21}%
  \BibitemOpen
  \bibfield  {author} {\bibinfo {author} {\bibfnamefont {M.}~\bibnamefont
  {Fruchart}}, \bibinfo {author} {\bibfnamefont {R.}~\bibnamefont {Hanai}},
  \bibinfo {author} {\bibfnamefont {P.~B.}\ \bibnamefont {Littlewood}}, \ and\
  \bibinfo {author} {\bibfnamefont {V.}~\bibnamefont {Vitelli}},\ }\href
  {\doibase 10.1038/s41586-021-03375-9} {\bibfield  {journal} {\bibinfo
  {journal} {Nature}\ }\textbf {\bibinfo {volume} {592}},\ \bibinfo {pages}
  {363} (\bibinfo {year} {2021})}\BibitemShut {NoStop}%
\bibitem [{\citenamefont {Rana}\ and\ \citenamefont
  {Golestanian}(2024)}]{rana24}%
  \BibitemOpen
  \bibfield  {author} {\bibinfo {author} {\bibfnamefont {N.}~\bibnamefont
  {Rana}}\ and\ \bibinfo {author} {\bibfnamefont {R.}~\bibnamefont
  {Golestanian}},\ }\href {\doibase 10.1103/PhysRevLett.133.078301} {\bibfield
  {journal} {\bibinfo  {journal} {Phys. Rev. Lett.}\ }\textbf {\bibinfo
  {volume} {133}},\ \bibinfo {pages} {078301} (\bibinfo {year}
  {2024})}\BibitemShut {NoStop}%
\bibitem [{\citenamefont {Guislain}\ and\ \citenamefont
  {Bertin}(2024)}]{guis24}%
  \BibitemOpen
  \bibfield  {author} {\bibinfo {author} {\bibfnamefont {L.}~\bibnamefont
  {Guislain}}\ and\ \bibinfo {author} {\bibfnamefont {E.}~\bibnamefont
  {Bertin}},\ }\href {\doibase 10.1088/1742-5468/ad72dc} {\bibfield  {journal}
  {\bibinfo  {journal} {J. Stat. Mech.}\ ,\ \bibinfo {pages} {093210}}
  (\bibinfo {year} {2024})}\BibitemShut {NoStop}%
\bibitem [{\citenamefont {Szolnoki}\ \emph {et~al.}(2014)\citenamefont
  {Szolnoki}, \citenamefont {Mobilia}, \citenamefont {Jiang}, \citenamefont
  {Szczesny}, \citenamefont {Rucklidge},\ and\ \citenamefont {Perc}}]{szol14}%
  \BibitemOpen
  \bibfield  {author} {\bibinfo {author} {\bibfnamefont {A.}~\bibnamefont
  {Szolnoki}}, \bibinfo {author} {\bibfnamefont {M.}~\bibnamefont {Mobilia}},
  \bibinfo {author} {\bibfnamefont {L.-L.}\ \bibnamefont {Jiang}}, \bibinfo
  {author} {\bibfnamefont {B.}~\bibnamefont {Szczesny}}, \bibinfo {author}
  {\bibfnamefont {A.~M.}\ \bibnamefont {Rucklidge}}, \ and\ \bibinfo {author}
  {\bibfnamefont {M.}~\bibnamefont {Perc}},\ }\href {\doibase
  10.1098/rsif.2014.0735} {\bibfield  {journal} {\bibinfo  {journal} {J. R.
  Soc. Interface}\ }\textbf {\bibinfo {volume} {11}},\ \bibinfo {pages}
  {20140735} (\bibinfo {year} {2014})}\BibitemShut {NoStop}%
\bibitem [{\citenamefont {Szab\'o}\ and\ \citenamefont
  {Szolnoki}(2002)}]{szab02}%
  \BibitemOpen
  \bibfield  {author} {\bibinfo {author} {\bibfnamefont {G.}~\bibnamefont
  {Szab\'o}}\ and\ \bibinfo {author} {\bibfnamefont {A.}~\bibnamefont
  {Szolnoki}},\ }\href {\doibase 10.1103/PhysRevE.65.036115} {\bibfield
  {journal} {\bibinfo  {journal} {Phys. Rev. E}\ }\textbf {\bibinfo {volume}
  {65}},\ \bibinfo {pages} {036115} (\bibinfo {year} {2002})}\BibitemShut
  {NoStop}%
\bibitem [{\citenamefont {Reichenbach}\ \emph {et~al.}(2007)\citenamefont
  {Reichenbach}, \citenamefont {Mobilia},\ and\ \citenamefont {Frey}}]{reic07}%
  \BibitemOpen
  \bibfield  {author} {\bibinfo {author} {\bibfnamefont {T.}~\bibnamefont
  {Reichenbach}}, \bibinfo {author} {\bibfnamefont {M.}~\bibnamefont
  {Mobilia}}, \ and\ \bibinfo {author} {\bibfnamefont {E.}~\bibnamefont
  {Frey}},\ }\href {\doibase 10.1038/nature06095} {\bibfield  {journal}
  {\bibinfo  {journal} {Nature}\ }\textbf {\bibinfo {volume} {448}},\ \bibinfo
  {pages} {1046} (\bibinfo {year} {2007})}\BibitemShut {NoStop}%
\bibitem [{\citenamefont {Szczesny}\ \emph {et~al.}(2013)\citenamefont
  {Szczesny}, \citenamefont {Mobilia},\ and\ \citenamefont
  {Rucklidge}}]{szcz13}%
  \BibitemOpen
  \bibfield  {author} {\bibinfo {author} {\bibfnamefont {B.}~\bibnamefont
  {Szczesny}}, \bibinfo {author} {\bibfnamefont {M.}~\bibnamefont {Mobilia}}, \
  and\ \bibinfo {author} {\bibfnamefont {A.~M.}\ \bibnamefont {Rucklidge}},\
  }\href {\doibase 10.1209/0295-5075/102/28012} {\bibfield  {journal} {\bibinfo
   {journal} {EPL}\ }\textbf {\bibinfo {volume} {102}},\ \bibinfo {pages}
  {28012} (\bibinfo {year} {2013})}\BibitemShut {NoStop}%
\bibitem [{\citenamefont {Kelsic}\ \emph {et~al.}(2015)\citenamefont {Kelsic},
  \citenamefont {Zhao}, \citenamefont {Vetsigian},\ and\ \citenamefont
  {Kishony}}]{kels15}%
  \BibitemOpen
  \bibfield  {author} {\bibinfo {author} {\bibfnamefont {E.~D.}\ \bibnamefont
  {Kelsic}}, \bibinfo {author} {\bibfnamefont {J.}~\bibnamefont {Zhao}},
  \bibinfo {author} {\bibfnamefont {K.}~\bibnamefont {Vetsigian}}, \ and\
  \bibinfo {author} {\bibfnamefont {R.}~\bibnamefont {Kishony}},\ }\href
  {\doibase 10.1038/nature14485} {\bibfield  {journal} {\bibinfo  {journal}
  {Nature}\ }\textbf {\bibinfo {volume} {521}},\ \bibinfo {pages} {516}
  (\bibinfo {year} {2015})}\BibitemShut {NoStop}%
\bibitem [{\citenamefont {Dobramysl}\ \emph {et~al.}(2018)\citenamefont
  {Dobramysl}, \citenamefont {Mobilia}, \citenamefont {Pleimling},\ and\
  \citenamefont {T\"auber}}]{dobr18}%
  \BibitemOpen
  \bibfield  {author} {\bibinfo {author} {\bibfnamefont {U.}~\bibnamefont
  {Dobramysl}}, \bibinfo {author} {\bibfnamefont {M.}~\bibnamefont {Mobilia}},
  \bibinfo {author} {\bibfnamefont {M.}~\bibnamefont {Pleimling}}, \ and\
  \bibinfo {author} {\bibfnamefont {U.~C.}\ \bibnamefont {T\"auber}},\ }\href
  {\doibase 10.1088/1751-8121/aa95c7} {\bibfield  {journal} {\bibinfo
  {journal} {J. Phys. A: Math. Theor.}\ }\textbf {\bibinfo {volume} {51}},\
  \bibinfo {pages} {063001} (\bibinfo {year} {2018})}\BibitemShut {NoStop}%
\bibitem [{\citenamefont {Szab\'o}\ and\ \citenamefont
  {Arial~Sznaider}(2004)}]{szab04}%
  \BibitemOpen
  \bibfield  {author} {\bibinfo {author} {\bibfnamefont {G.}~\bibnamefont
  {Szab\'o}}\ and\ \bibinfo {author} {\bibfnamefont {G.}~\bibnamefont
  {Arial~Sznaider}},\ }\href {\doibase 10.1103/PhysRevE.69.031911} {\bibfield
  {journal} {\bibinfo  {journal} {Phys. Rev. E}\ }\textbf {\bibinfo {volume}
  {69}},\ \bibinfo {pages} {031911} (\bibinfo {year} {2004})}\BibitemShut
  {NoStop}%
\bibitem [{\citenamefont {Szab\'o}\ \emph {et~al.}(2008)\citenamefont
  {Szab\'o}, \citenamefont {Szolnoki},\ and\ \citenamefont {Borsos}}]{szab08}%
  \BibitemOpen
  \bibfield  {author} {\bibinfo {author} {\bibfnamefont {G.}~\bibnamefont
  {Szab\'o}}, \bibinfo {author} {\bibfnamefont {A.}~\bibnamefont {Szolnoki}}, \
  and\ \bibinfo {author} {\bibfnamefont {I.}~\bibnamefont {Borsos}},\ }\href
  {\doibase 10.1103/PhysRevE.77.041919} {\bibfield  {journal} {\bibinfo
  {journal} {Phys. Rev. E}\ }\textbf {\bibinfo {volume} {77}},\ \bibinfo
  {pages} {041919} (\bibinfo {year} {2008})}\BibitemShut {NoStop}%
\bibitem [{\citenamefont {Roman}\ \emph {et~al.}(2012)\citenamefont {Roman},
  \citenamefont {Konrad},\ and\ \citenamefont {Pleimling}}]{roma12}%
  \BibitemOpen
  \bibfield  {author} {\bibinfo {author} {\bibfnamefont {A.}~\bibnamefont
  {Roman}}, \bibinfo {author} {\bibfnamefont {D.}~\bibnamefont {Konrad}}, \
  and\ \bibinfo {author} {\bibfnamefont {M.}~\bibnamefont {Pleimling}},\ }\href
  {\doibase 10.1088/1742-5468/2012/07/P07014} {\bibfield  {journal} {\bibinfo
  {journal} {J. Stat. Mech.}\ ,\ \bibinfo {pages} {P07014}} (\bibinfo {year}
  {2012})}\BibitemShut {NoStop}%
\bibitem [{\citenamefont {Rulquin}\ and\ \citenamefont
  {Arenzon}(2014)}]{rulq14}%
  \BibitemOpen
  \bibfield  {author} {\bibinfo {author} {\bibfnamefont {C.}~\bibnamefont
  {Rulquin}}\ and\ \bibinfo {author} {\bibfnamefont {J.~J.}\ \bibnamefont
  {Arenzon}},\ }\href {\doibase 10.1103/PhysRevE.89.032133} {\bibfield
  {journal} {\bibinfo  {journal} {Phys. Rev. E}\ }\textbf {\bibinfo {volume}
  {89}},\ \bibinfo {pages} {032133} (\bibinfo {year} {2014})}\BibitemShut
  {NoStop}%
\bibitem [{\citenamefont {Bazeia}\ \emph {et~al.}(2019)\citenamefont {Bazeia},
  \citenamefont {de~Oliveira},\ and\ \citenamefont {Szolnoki}}]{baze19}%
  \BibitemOpen
  \bibfield  {author} {\bibinfo {author} {\bibfnamefont {D.}~\bibnamefont
  {Bazeia}}, \bibinfo {author} {\bibfnamefont {B.~F.}\ \bibnamefont
  {de~Oliveira}}, \ and\ \bibinfo {author} {\bibfnamefont {A.}~\bibnamefont
  {Szolnoki}},\ }\href {\doibase 10.1103/PhysRevE.99.052408} {\bibfield
  {journal} {\bibinfo  {journal} {Phys. Rev. E}\ }\textbf {\bibinfo {volume}
  {99}},\ \bibinfo {pages} {052408} (\bibinfo {year} {2019})}\BibitemShut
  {NoStop}%
\bibitem [{\citenamefont {Zhong}\ \emph {et~al.}(2022)\citenamefont {Zhong},
  \citenamefont {Zhang}, \citenamefont {Li}, \citenamefont {Dai},\ and\
  \citenamefont {Yang}}]{zhon22}%
  \BibitemOpen
  \bibfield  {author} {\bibinfo {author} {\bibfnamefont {L.}~\bibnamefont
  {Zhong}}, \bibinfo {author} {\bibfnamefont {L.}~\bibnamefont {Zhang}},
  \bibinfo {author} {\bibfnamefont {H.}~\bibnamefont {Li}}, \bibinfo {author}
  {\bibfnamefont {Q.}~\bibnamefont {Dai}}, \ and\ \bibinfo {author}
  {\bibfnamefont {J.}~\bibnamefont {Yang}},\ }\href {\doibase
  10.1016/j.chaos.2022.111806} {\bibfield  {journal} {\bibinfo  {journal}
  {Chaos Soliton. Fract.}\ }\textbf {\bibinfo {volume} {156}},\ \bibinfo
  {pages} {111806} (\bibinfo {year} {2022})}\BibitemShut {NoStop}%
\bibitem [{\citenamefont {Yang}\ and\ \citenamefont {Park}(2023)}]{yang23}%
  \BibitemOpen
  \bibfield  {author} {\bibinfo {author} {\bibfnamefont {R.~K.}\ \bibnamefont
  {Yang}}\ and\ \bibinfo {author} {\bibfnamefont {J.}~\bibnamefont {Park}},\
  }\href {\doibase 10.1016/j.chaos.2023.113949} {\bibfield  {journal} {\bibinfo
   {journal} {Chaos Soliton. Fract.}\ }\textbf {\bibinfo {volume} {175}},\
  \bibinfo {pages} {113949} (\bibinfo {year} {2023})}\BibitemShut {NoStop}%
\bibitem [{\citenamefont {Szolnoki}\ and\ \citenamefont {Chen}(2023)}]{szol23}%
  \BibitemOpen
  \bibfield  {author} {\bibinfo {author} {\bibfnamefont {A.}~\bibnamefont
  {Szolnoki}}\ and\ \bibinfo {author} {\bibfnamefont {X.}~\bibnamefont
  {Chen}},\ }\href {\doibase 10.1038/s41598-023-35746-9} {\bibfield  {journal}
  {\bibinfo  {journal} {Sci. Rep.}\ }\textbf {\bibinfo {volume} {13}},\
  \bibinfo {pages} {8472} (\bibinfo {year} {2023})}\BibitemShut {NoStop}%
\bibitem [{\citenamefont {Noguchi}\ \emph {et~al.}(2024)\citenamefont
  {Noguchi}, \citenamefont {{van Wijland}},\ and\ \citenamefont
  {Fournier}}]{nogu24a}%
  \BibitemOpen
  \bibfield  {author} {\bibinfo {author} {\bibfnamefont {H.}~\bibnamefont
  {Noguchi}}, \bibinfo {author} {\bibfnamefont {F.}~\bibnamefont {{van
  Wijland}}}, \ and\ \bibinfo {author} {\bibfnamefont {J.-B.}\ \bibnamefont
  {Fournier}},\ }\href {\doibase 10.1063/5.0221050} {\bibfield  {journal}
  {\bibinfo  {journal} {J. Chem. Phys.}\ }\textbf {\bibinfo {volume} {161}},\
  \bibinfo {pages} {025101} (\bibinfo {year} {2024})}\BibitemShut {NoStop}%
\bibitem [{\citenamefont {Noguchi}\ and\ \citenamefont
  {Fournier}(2024)}]{nogu24b}%
  \BibitemOpen
  \bibfield  {author} {\bibinfo {author} {\bibfnamefont {H.}~\bibnamefont
  {Noguchi}}\ and\ \bibinfo {author} {\bibfnamefont {J.-B.}\ \bibnamefont
  {Fournier}},\ }\href {\doibase 10.1088/1367-2630/ad7dac} {\bibfield
  {journal} {\bibinfo  {journal} {New J. Phys.}\ }\textbf {\bibinfo {volume}
  {26}},\ \bibinfo {pages} {093043} (\bibinfo {year} {2024})}\BibitemShut
  {NoStop}%
\bibitem [{\citenamefont {Noguchi}(2025{\natexlab{b}})}]{nogu25}%
  \BibitemOpen
  \bibfield  {author} {\bibinfo {author} {\bibfnamefont {H.}~\bibnamefont
  {Noguchi}},\ }\href {\doibase 10.1038/s41598-024-84819-w} {\bibfield
  {journal} {\bibinfo  {journal} {Sci. Rep.}\ }\textbf {\bibinfo {volume}
  {15}},\ \bibinfo {pages} {674} (\bibinfo {year}
  {2025}{\natexlab{b}})}\BibitemShut {NoStop}%
\bibitem [{\citenamefont {Noguchi}(2025{\natexlab{c}})}]{nogu25a}%
  \BibitemOpen
  \bibfield  {author} {\bibinfo {author} {\bibfnamefont {H.}~\bibnamefont
  {Noguchi}},\ }\href {\doibase 10.1039/D4SM01277A} {\bibfield  {journal}
  {\bibinfo  {journal} {Soft Matter}\ }\textbf {\bibinfo {volume} {21}},\
  \bibinfo {pages} {1113} (\bibinfo {year} {2025}{\natexlab{c}})}\BibitemShut
  {NoStop}%
\bibitem [{\citenamefont {Noguchi}(2025{\natexlab{d}})}]{nogu25b}%
  \BibitemOpen
  \bibfield  {author} {\bibinfo {author} {\bibfnamefont {H.}~\bibnamefont
  {Noguchi}},\ }\href {\doibase 10.1103/w1fg-6qmv} {\bibfield  {journal}
  {\bibinfo  {journal} {Phys. Rev. Res.}\ }\textbf {\bibinfo {volume} {7}},\
  \bibinfo {pages} {033243} (\bibinfo {year} {2025}{\natexlab{d}})}\BibitemShut
  {NoStop}%
\bibitem [{\citenamefont {Noguchi}()}]{nogu25c}%
  \BibitemOpen
  \bibfield  {author} {\bibinfo {author} {\bibfnamefont {H.}~\bibnamefont
  {Noguchi}},\ }\href@noop {} {}\Eprint {http://arxiv.org/abs/arXiv:2509.17408}
  {arXiv:2509.17408} \BibitemShut {NoStop}%
\bibitem [{\citenamefont {Risler}\ \emph {et~al.}(2004)\citenamefont {Risler},
  \citenamefont {Prost},\ and\ \citenamefont {J\"ulicher}}]{risl04}%
  \BibitemOpen
  \bibfield  {author} {\bibinfo {author} {\bibfnamefont {T.}~\bibnamefont
  {Risler}}, \bibinfo {author} {\bibfnamefont {J.}~\bibnamefont {Prost}}, \
  and\ \bibinfo {author} {\bibfnamefont {F.}~\bibnamefont {J\"ulicher}},\
  }\href {\doibase 10.1103/PhysRevLett.93.175702} {\bibfield  {journal}
  {\bibinfo  {journal} {Phys. Rev. Lett.}\ }\textbf {\bibinfo {volume} {93}},\
  \bibinfo {pages} {175702} (\bibinfo {year} {2004})}\BibitemShut {NoStop}%
\bibitem [{\citenamefont {Wood}\ \emph {et~al.}(2006)\citenamefont {Wood},
  \citenamefont {Van~den Broeck}, \citenamefont {Kawai},\ and\ \citenamefont
  {Lindenberg}}]{wood06}%
  \BibitemOpen
  \bibfield  {author} {\bibinfo {author} {\bibfnamefont {K.}~\bibnamefont
  {Wood}}, \bibinfo {author} {\bibfnamefont {C.}~\bibnamefont {Van~den
  Broeck}}, \bibinfo {author} {\bibfnamefont {R.}~\bibnamefont {Kawai}}, \ and\
  \bibinfo {author} {\bibfnamefont {K.}~\bibnamefont {Lindenberg}},\ }\href
  {\doibase 10.1103/PhysRevLett.96.145701} {\bibfield  {journal} {\bibinfo
  {journal} {Phys. Rev. Lett.}\ }\textbf {\bibinfo {volume} {96}},\ \bibinfo
  {pages} {145701} (\bibinfo {year} {2006})}\BibitemShut {NoStop}%
\bibitem [{\citenamefont {Avni}\ \emph {et~al.}(2025)\citenamefont {Avni},
  \citenamefont {Fruchart}, \citenamefont {Martin}, \citenamefont {Seara},\
  and\ \citenamefont {Vitelli}}]{avni25}%
  \BibitemOpen
  \bibfield  {author} {\bibinfo {author} {\bibfnamefont {Y.}~\bibnamefont
  {Avni}}, \bibinfo {author} {\bibfnamefont {M.}~\bibnamefont {Fruchart}},
  \bibinfo {author} {\bibfnamefont {D.}~\bibnamefont {Martin}}, \bibinfo
  {author} {\bibfnamefont {D.}~\bibnamefont {Seara}}, \ and\ \bibinfo {author}
  {\bibfnamefont {V.}~\bibnamefont {Vitelli}},\ }\href {\doibase
  10.1103/PhysRevLett.134.117103} {\bibfield  {journal} {\bibinfo  {journal}
  {Phys. Rev. Lett.}\ }\textbf {\bibinfo {volume} {134}},\ \bibinfo {pages}
  {117103} (\bibinfo {year} {2025})}\BibitemShut {NoStop}%
\bibitem [{\citenamefont {Herpich}\ \emph {et~al.}(2018)\citenamefont
  {Herpich}, \citenamefont {Thingna},\ and\ \citenamefont {Esposito}}]{herp18}%
  \BibitemOpen
  \bibfield  {author} {\bibinfo {author} {\bibfnamefont {T.}~\bibnamefont
  {Herpich}}, \bibinfo {author} {\bibfnamefont {J.}~\bibnamefont {Thingna}}, \
  and\ \bibinfo {author} {\bibfnamefont {M.}~\bibnamefont {Esposito}},\ }\href
  {\doibase 10.1103/PhysRevX.8.031056} {\bibfield  {journal} {\bibinfo
  {journal} {Phys. Rev. X}\ }\textbf {\bibinfo {volume} {8}},\ \bibinfo {pages}
  {031056} (\bibinfo {year} {2018})}\BibitemShut {NoStop}%
\bibitem [{\citenamefont {Meibohm}\ and\ \citenamefont
  {Esposito}(2024)}]{meib24}%
  \BibitemOpen
  \bibfield  {author} {\bibinfo {author} {\bibfnamefont {J.}~\bibnamefont
  {Meibohm}}\ and\ \bibinfo {author} {\bibfnamefont {M.}~\bibnamefont
  {Esposito}},\ }\href {\doibase 10.1103/PhysRevE.110.044114} {\bibfield
  {journal} {\bibinfo  {journal} {Phys. Rev. E}\ }\textbf {\bibinfo {volume}
  {110}},\ \bibinfo {pages} {044114} (\bibinfo {year} {2024})}\BibitemShut
  {NoStop}%
\bibitem [{\citenamefont {Ptaszy\ifmmode~\acute{n}\else \'{n}\fi{}ski}\ and\
  \citenamefont {Esposito}(2025)}]{ptas25}%
  \BibitemOpen
  \bibfield  {author} {\bibinfo {author} {\bibfnamefont {K.}~\bibnamefont
  {Ptaszy\ifmmode~\acute{n}\else \'{n}\fi{}ski}}\ and\ \bibinfo {author}
  {\bibfnamefont {M.}~\bibnamefont {Esposito}},\ }\href {\doibase
  10.1103/j21x-hrsq} {\bibfield  {journal} {\bibinfo  {journal} {Phys. Rev.
  Lett.}\ }\textbf {\bibinfo {volume} {135}},\ \bibinfo {pages} {057401}
  (\bibinfo {year} {2025})}\BibitemShut {NoStop}%
\bibitem [{\citenamefont {Nov{\'a}k}\ and\ \citenamefont
  {Tyson}(2008)}]{nova08}%
  \BibitemOpen
  \bibfield  {author} {\bibinfo {author} {\bibfnamefont {B.}~\bibnamefont
  {Nov{\'a}k}}\ and\ \bibinfo {author} {\bibfnamefont {J.~J.}\ \bibnamefont
  {Tyson}},\ }\href {\doibase 10.1038/nrm2530} {\bibfield  {journal} {\bibinfo
  {journal} {Nat. Rev. Mol. Cell Biol.}\ }\textbf {\bibinfo {volume} {9}},\
  \bibinfo {pages} {981} (\bibinfo {year} {2008})}\BibitemShut {NoStop}%
\bibitem [{\citenamefont {Arenas}\ \emph {et~al.}(2008)\citenamefont {Arenas},
  \citenamefont {D\'{i}az-Guilera}, \citenamefont {Kurths}, \citenamefont
  {Moreno},\ and\ \citenamefont {Zhou}}]{aren08}%
  \BibitemOpen
  \bibfield  {author} {\bibinfo {author} {\bibfnamefont {A.}~\bibnamefont
  {Arenas}}, \bibinfo {author} {\bibfnamefont {A.}~\bibnamefont
  {D\'{i}az-Guilera}}, \bibinfo {author} {\bibfnamefont {J.}~\bibnamefont
  {Kurths}}, \bibinfo {author} {\bibfnamefont {Y.}~\bibnamefont {Moreno}}, \
  and\ \bibinfo {author} {\bibfnamefont {C.}~\bibnamefont {Zhou}},\ }\href
  {\doibase 10.1016/j.physrep.2008.09.002} {\bibfield  {journal} {\bibinfo
  {journal} {Phys. Rep.}\ }\textbf {\bibinfo {volume} {469}},\ \bibinfo {pages}
  {93} (\bibinfo {year} {2008})}\BibitemShut {NoStop}%
\bibitem [{\citenamefont {Artime}\ \emph {et~al.}(2024)\citenamefont {Artime},
  \citenamefont {Grassia}, \citenamefont {Domenico}, \citenamefont {Gleeson},
  \citenamefont {Makse}, \citenamefont {Mangioni}, \citenamefont {Perc},\ and\
  \citenamefont {Radicchi}}]{arti24}%
  \BibitemOpen
  \bibfield  {author} {\bibinfo {author} {\bibfnamefont {O.}~\bibnamefont
  {Artime}}, \bibinfo {author} {\bibfnamefont {M.}~\bibnamefont {Grassia}},
  \bibinfo {author} {\bibfnamefont {M.~D.}\ \bibnamefont {Domenico}}, \bibinfo
  {author} {\bibfnamefont {J.~P.}\ \bibnamefont {Gleeson}}, \bibinfo {author}
  {\bibfnamefont {H.~A.}\ \bibnamefont {Makse}}, \bibinfo {author}
  {\bibfnamefont {G.}~\bibnamefont {Mangioni}}, \bibinfo {author}
  {\bibfnamefont {M.}~\bibnamefont {Perc}}, \ and\ \bibinfo {author}
  {\bibfnamefont {F.}~\bibnamefont {Radicchi}},\ }\href {\doibase
  10.1038/s42254-023-00676-y} {\bibfield  {journal} {\bibinfo  {journal} {Nat.
  Rev. Phys.}\ }\textbf {\bibinfo {volume} {6}},\ \bibinfo {pages} {114}
  (\bibinfo {year} {2024})}\BibitemShut {NoStop}%
\bibitem [{\citenamefont {Wu}(1982)}]{wu82}%
  \BibitemOpen
  \bibfield  {author} {\bibinfo {author} {\bibfnamefont {F.~Y.}\ \bibnamefont
  {Wu}},\ }\href {\doibase 10.1103/RevModPhys.54.235} {\bibfield  {journal}
  {\bibinfo  {journal} {Rev. Mod. Phys.}\ }\textbf {\bibinfo {volume} {54}},\
  \bibinfo {pages} {235} (\bibinfo {year} {1982})}\BibitemShut {NoStop}%
\bibitem [{\citenamefont {Potts}(1952)}]{pott52}%
  \BibitemOpen
  \bibfield  {author} {\bibinfo {author} {\bibfnamefont {R.~B.}\ \bibnamefont
  {Potts}},\ }\href {\doibase 10.1017/S0305004100027419} {\bibfield  {journal}
  {\bibinfo  {journal} {Proc. Camb. Phil. Soc.}\ }\textbf {\bibinfo {volume}
  {48}},\ \bibinfo {pages} {106} (\bibinfo {year} {1952})}\BibitemShut
  {NoStop}%
\bibitem [{\citenamefont {Ertl}(2008)}]{ertl08}%
  \BibitemOpen
  \bibfield  {author} {\bibinfo {author} {\bibfnamefont {G.}~\bibnamefont
  {Ertl}},\ }\href {\doibase 10.1002/anie.200800480} {\bibfield  {journal}
  {\bibinfo  {journal} {Angew. Chem. Int. Ed.}\ }\textbf {\bibinfo {volume}
  {47}},\ \bibinfo {pages} {3524} (\bibinfo {year} {2008})}\BibitemShut
  {NoStop}%
\bibitem [{\citenamefont {Br{\"a}r}\ \emph {et~al.}(1994)\citenamefont
  {Br{\"a}r}, \citenamefont {Gottschalk}, \citenamefont {Eiswirth},\ and\
  \citenamefont {Ertl}}]{bar94}%
  \BibitemOpen
  \bibfield  {author} {\bibinfo {author} {\bibfnamefont {M.}~\bibnamefont
  {Br{\"a}r}}, \bibinfo {author} {\bibfnamefont {N.}~\bibnamefont
  {Gottschalk}}, \bibinfo {author} {\bibfnamefont {M.}~\bibnamefont
  {Eiswirth}}, \ and\ \bibinfo {author} {\bibfnamefont {G.}~\bibnamefont
  {Ertl}},\ }\href {\doibase 10.1063/1.466650} {\bibfield  {journal} {\bibinfo
  {journal} {J. Chem. Phys.}\ }\textbf {\bibinfo {volume} {100}},\ \bibinfo
  {pages} {1202} (\bibinfo {year} {1994})}\BibitemShut {NoStop}%
\bibitem [{\citenamefont {Gorodetskii}\ \emph {et~al.}(1994)\citenamefont
  {Gorodetskii}, \citenamefont {Lauterbach}, \citenamefont {Rotermund},
  \citenamefont {Block},\ and\ \citenamefont {Ertl}}]{goro94}%
  \BibitemOpen
  \bibfield  {author} {\bibinfo {author} {\bibfnamefont {V.}~\bibnamefont
  {Gorodetskii}}, \bibinfo {author} {\bibfnamefont {J.}~\bibnamefont
  {Lauterbach}}, \bibinfo {author} {\bibfnamefont {H.-H.}\ \bibnamefont
  {Rotermund}}, \bibinfo {author} {\bibfnamefont {J.~H.}\ \bibnamefont
  {Block}}, \ and\ \bibinfo {author} {\bibfnamefont {G.}~\bibnamefont {Ertl}},\
  }\href {\doibase 10.1038/370276a0} {\bibfield  {journal} {\bibinfo  {journal}
  {Nature}\ }\textbf {\bibinfo {volume} {370}},\ \bibinfo {pages} {276}
  (\bibinfo {year} {1994})}\BibitemShut {NoStop}%
\bibitem [{\citenamefont {Barroo}\ \emph {et~al.}(2020)\citenamefont {Barroo},
  \citenamefont {Wang}, \citenamefont {Schlr{\"o}gl},\ and\ \citenamefont
  {Willinger}}]{barr20}%
  \BibitemOpen
  \bibfield  {author} {\bibinfo {author} {\bibfnamefont {C.}~\bibnamefont
  {Barroo}}, \bibinfo {author} {\bibfnamefont {Z.-J.}\ \bibnamefont {Wang}},
  \bibinfo {author} {\bibfnamefont {R.}~\bibnamefont {Schlr{\"o}gl}}, \ and\
  \bibinfo {author} {\bibfnamefont {M.-G.}\ \bibnamefont {Willinger}},\ }\href
  {\doibase 10.1038/s41929-019-0395-3} {\bibfield  {journal} {\bibinfo
  {journal} {Nat. Catal.}\ }\textbf {\bibinfo {volume} {3}},\ \bibinfo {pages}
  {30} (\bibinfo {year} {2020})}\BibitemShut {NoStop}%
\bibitem [{\citenamefont {Zeininger}\ and\ \citenamefont
  {et~al.}(2022)}]{zein22}%
  \BibitemOpen
  \bibfield  {author} {\bibinfo {author} {\bibfnamefont {J.}~\bibnamefont
  {Zeininger}}\ and\ \bibinfo {author} {\bibnamefont {et~al.}},\ }\href
  {\doibase 10.1021/acscatal.2c03692} {\bibfield  {journal} {\bibinfo
  {journal} {ACS Catal.}\ }\textbf {\bibinfo {volume} {12}},\ \bibinfo {pages}
  {11974} (\bibinfo {year} {2022})}\BibitemShut {NoStop}%
\bibitem [{\citenamefont {Tabe}\ and\ \citenamefont {Yokoyama}(2003)}]{tabe03}%
  \BibitemOpen
  \bibfield  {author} {\bibinfo {author} {\bibfnamefont {Y.}~\bibnamefont
  {Tabe}}\ and\ \bibinfo {author} {\bibfnamefont {H.}~\bibnamefont
  {Yokoyama}},\ }\href {\doibase 10.1038/nmat1017} {\bibfield  {journal}
  {\bibinfo  {journal} {Nat. Mater.}\ }\textbf {\bibinfo {volume} {2}},\
  \bibinfo {pages} {806} (\bibinfo {year} {2003})}\BibitemShut {NoStop}%
\bibitem [{\citenamefont {Miele}\ \emph {et~al.}(2020)\citenamefont {Miele},
  \citenamefont {Medveczky}, \citenamefont {Holl{\'o}}, \citenamefont {Tegze},
  \citenamefont {Der{\'e}nyi}, \citenamefont {H{\'o}rv{\"o}lgyi}, \citenamefont
  {Altamura}, \citenamefont {Lagzi},\ and\ \citenamefont {Rossi}}]{miel20}%
  \BibitemOpen
  \bibfield  {author} {\bibinfo {author} {\bibfnamefont {Y.}~\bibnamefont
  {Miele}}, \bibinfo {author} {\bibfnamefont {Z.}~\bibnamefont {Medveczky}},
  \bibinfo {author} {\bibfnamefont {G.}~\bibnamefont {Holl{\'o}}}, \bibinfo
  {author} {\bibfnamefont {B.}~\bibnamefont {Tegze}}, \bibinfo {author}
  {\bibfnamefont {I.}~\bibnamefont {Der{\'e}nyi}}, \bibinfo {author}
  {\bibfnamefont {Z.}~\bibnamefont {H{\'o}rv{\"o}lgyi}}, \bibinfo {author}
  {\bibfnamefont {E.}~\bibnamefont {Altamura}}, \bibinfo {author}
  {\bibfnamefont {I.}~\bibnamefont {Lagzi}}, \ and\ \bibinfo {author}
  {\bibfnamefont {F.}~\bibnamefont {Rossi}},\ }\href {\doibase
  10.1039/c9sc05195c} {\bibfield  {journal} {\bibinfo  {journal} {Chem. Sci.}\
  }\textbf {\bibinfo {volume} {11}},\ \bibinfo {pages} {3228} (\bibinfo {year}
  {2020})}\BibitemShut {NoStop}%
\bibitem [{\citenamefont {Holl{\'o}}\ \emph {et~al.}(2021)\citenamefont
  {Holl{\'o}}, \citenamefont {Miele}, \citenamefont {Rossi},\ and\
  \citenamefont {Lagzi}}]{holl21}%
  \BibitemOpen
  \bibfield  {author} {\bibinfo {author} {\bibfnamefont {G.}~\bibnamefont
  {Holl{\'o}}}, \bibinfo {author} {\bibfnamefont {Y.}~\bibnamefont {Miele}},
  \bibinfo {author} {\bibfnamefont {F.}~\bibnamefont {Rossi}}, \ and\ \bibinfo
  {author} {\bibfnamefont {I.}~\bibnamefont {Lagzi}},\ }\href {\doibase
  10.1039/d0cp05952h} {\bibfield  {journal} {\bibinfo  {journal} {Phys. Chem.
  Chem. Phys.}\ }\textbf {\bibinfo {volume} {23}},\ \bibinfo {pages} {4262}
  (\bibinfo {year} {2021})}\BibitemShut {NoStop}%
\bibitem [{\citenamefont {Noguchi}(2023)}]{nogu23}%
  \BibitemOpen
  \bibfield  {author} {\bibinfo {author} {\bibfnamefont {H.}~\bibnamefont
  {Noguchi}},\ }\href {\doibase 10.1039/d2sm01536f} {\bibfield  {journal}
  {\bibinfo  {journal} {Soft Matter}\ }\textbf {\bibinfo {volume} {19}},\
  \bibinfo {pages} {679} (\bibinfo {year} {2023})}\BibitemShut {NoStop}%
\bibitem [{\citenamefont {Baxter}(1973)}]{baxt73}%
  \BibitemOpen
  \bibfield  {author} {\bibinfo {author} {\bibfnamefont {R.~J.}\ \bibnamefont
  {Baxter}},\ }\href@noop {} {\bibfield  {journal} {\bibinfo  {journal} {J.
  Phys. C: Solid State Phys.}\ }\textbf {\bibinfo {volume} {6}},\ \bibinfo
  {pages} {L445} (\bibinfo {year} {1973})}\BibitemShut {NoStop}%
\end{thebibliography}

%

\end{document}